# THE FEASIBILITY OF NEAR-FIELD ODR BEAM-SIZE MONITORING AT 23 GEV AT FACET

A.H. Lumpkin*, C.-Y. Yao (ANL), M. Hogan (SLAC), and P. Muggli (USC)
*Fermilab, Batavia, IL U.S.A. 60510

## Abstract

Extension of near-field optical diffraction radiation (ODR) imaging to the 23 GeV beams at the proposed FACET facility at SLAC has been evaluated. The beam-size sensitivity at the 10- to 20-µm sigma level based on a simple model will be reported. Polarization effects are also seen to be important and will be discussed. The comparisons to previous experimental results and the modeling results indicate sufficient feasibility for planning of the experiments in the coming year.

## INTRODUCTION

The trend for high-average power, high-charge density, and/or low-emittance beams in present and future accelerator facilities has resulted in continued interest in the development of techniques for beam characterizations in a non-intercepting (NI) manner. One candidate for assessing the transverse size is optical diffraction radiation (ODR), which is emitted when a charged-particle beam passes near the edge of a conducting surface or through an aperture in the surface [1,2]. Most investigations have involved imaging at infinity (or far field) to determine the angular distribution pattern changes with beam size [3-5]. A new paradigm for a NI beam-size monitor based on imaging ODR in the near field (focus at the object) has been developed over the last few years [6-9]. So far this technique has been used for beam sizes larger than 100 µm (sigma) and at beam energies of 7 GeV (APS/ANL), 4.5 GeV (CEBAF/JLAB), and 0.9 GeV (FLASH/DESY) as summarized in Fig. 1.

Due to the radial polarization of the radiation mechanism, there are actually analysis methods for both the perpendicular and parallel ODR polarization components that exhibit beam size sensitivity in a simple model [6]. Determining beam size sensitivity at the challenging 10- to 20-µm level and at 23 GeV as found at FACET [10] would significantly extend the parameter space over which this ODR technique is applicable and would be relevant to anticipated CEBAF upgrade and proposed International Linear Collider (ILC) parameters [11]. The 9-mA ILC test facility at Fermilab [12] is also indicated in Fig.1. Depending on the results of the tests, an ODR imaging station might provide a NI beam monitor as well in support of the proposed plasma wakefield accelerator (PWFA) experiments at FACET.

-------------------------------
*Work supported by U.S. Department of Energy, Office of Science, Office of High Energy Physics, under Contract No. DE-AC02-06CH11357.

## INITIAL ODR IMAGING EXAMPLE

As a basic early example of the technique, we show the APS/ANL results at 7 GeV with a 3.3 nC of charge in a single micropulse [6]. The beam was extracted from the booster synchrotron and directed down a local beam dump line. A schematic of the setup is shown in Fig. 2. An rf BPM just before the station provided beam vertical position monitoring during scans of the impact parameter (IP) and the corresponding loss monitor signals from the downstream Cherenkov detector. The beam size was 1375 µm in x by 200 µm in y. This flat beam allowed us to move the screen edge 1.25 mm away vertically (6 sigma-y), but still obtain adequate ODR signal for a standard CCD analog camera. Sample images are shown in Fig. 3. Eight times more charge was used for the ODR image than the OTR image in this early demonstration. Since $\gamma\lambda/2\pi$ was 1.4 mm for a nominal 0.628 µm wavelength, we satisfied the preferred ODR impact parameter regime ($\gamma\lambda/2\pi \sim$ IP). In addition, we had reasonable sensitivity to the beam size since the ODR image profile was only about 25% larger than the 1375 µm OTR beam size (assumed to be close to the actual size). Initial polarization tests also showed that the perpendicularly polarized ODR component was narrower than the total ODR image profile.

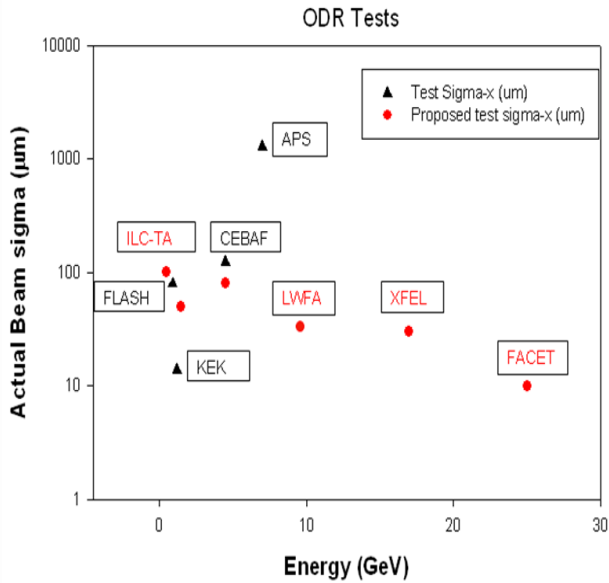

Figure 1: Summary chart for beam size vs. beam energy of the ODR tests done to date (black triangles) and proposed (red circles) with either far-field or near-field imaging. The near-field technique has been utilized at APS, CEBAF, and FLASH, and the proposed FACET parameters are a clear extension of possible applicability into future accelerators.

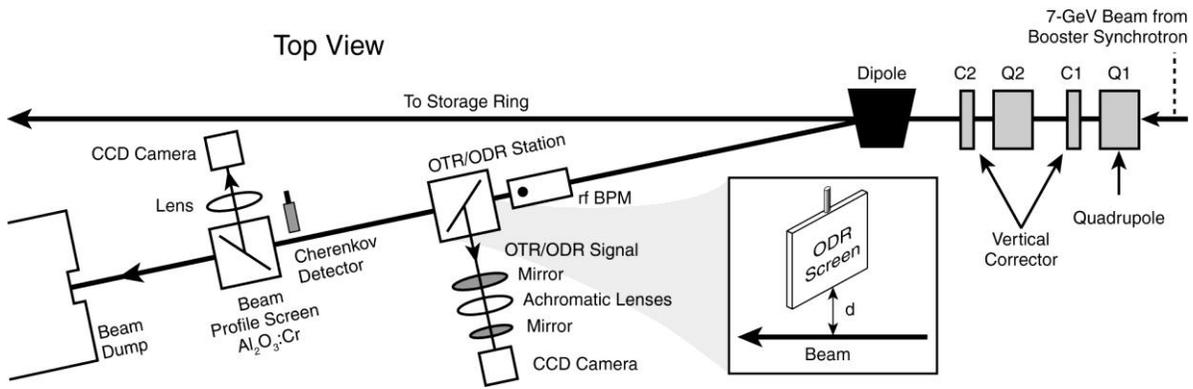

Figure. 2: Schematic of the OTR/ODR imaging station on the APS transport line. A beam of 7 GeV energy was extracted from the booster synchrotron with ~3 nC in a single micropulse. (From ref. 6).

The case for further tests includes the projected beam sizes that one might encounter in the ILC. With its projected high-average current, nonintercepting techniques are a requirement. An example parameter comparison list is provided in Table I. For ILC, both the 5 GeV and 250 GeV numbers are shown [10]. FACET numbers on beam size are ILC-like except for the smallest vertical size at 250 GeV. Successful imaging of the small beam sizes at high gamma at FACET would immediately extend the demonstrated parameter space. Figure 1 shows a summary chart of this aspect with the 10-µm beam size at high gamma being the challenging target for non-intercepting diagnostics. An ODR converter screen inserted by a stepper motor actuator is foreseen at FACET.

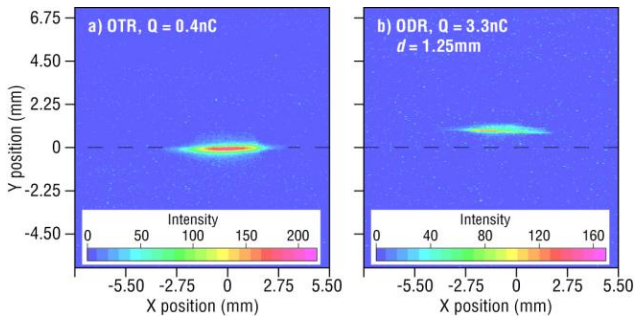

Figure 3: Early APS results showing a) the OTR image of the beam and b) the ODR image with impact parameter of 1.25 mm. The horizontal dashed line indicates the beam center position. Note the screen is inserted with a) its edge 4 mm below the centerline and b) 1.25 mm above the centerline. The ODR light comes from the screen and its induced surface currents. (From ref. 6).

## INITIAL SIMULATIONS

Some examples of early simulations are shown in this section. The basic measurement is the ODR image x profile. In Fig. 4a, the calculated vertical/perpendicular polarized component's x profile is shown for an impact parameter of 50 µm ($\gamma\lambda/2\pi$~5 mm). The actual input horizontal beam sizes were 10, 20, 35, 50, and 100 µm. Changes in the ODR profile size are seen that indicate sensitivity to the beam size change. These were evaluated at a wavelength of 0.8 µm and 25 GeV. Figure 4b summarizes the two impact-parameter cases. One sees that the 50-µm impact parameter has more sensitivity than that of the 100-µm case at the smallest beam sizes.

In addition, the parallel polarization component has a double-lobed structure in x for some conditions, and the depth of the valley between the lobes is calculated to be sensitive to horizontal beam size as seen in Fig. 5 for IPs of 100 and 50 µm. In the IP=50-µm case the horizontal component is about 3-4 times weaker in intensity than the vertical. The camera sensitivity will thus be an important factor in taking advantage of the parallel component signal. A scientific CMOS camera is being considered.

Table I. Comparison of beam parameters at existing and proposed facilities for ODR imaging investigations.

| Parameter | APS | CEBAF | ILC | FACET |
|---|---|---|---|---|
| Energy (GeV) | 7 | 1-5 | 5,250 | 23 |
| X Beam size (µm) | 1300 | 100-150 | 300,30 | 10 |
| Y Beam size (µm) | 200 | 100-150 | 15,2 | 10 |
| Current (nA) | 6 | 100,000 | 45,000 | 30 |
| Charge per 33 ms (nC) | 3 | 3,000 | 9,000 | 3 |

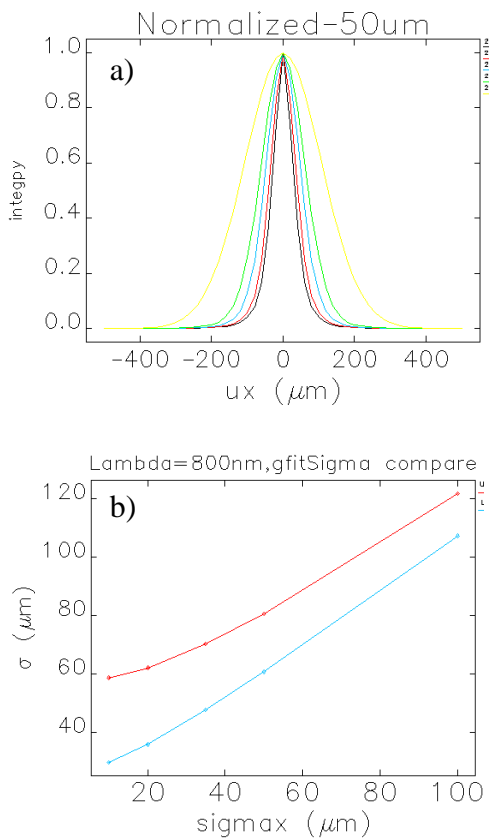

Figure 4: Numerical simulations of a) the ODR vertically polarized component changes with horizontal beam size (10-, 20-, 35-, 50-, and 100-μm cases are the black, red, blue, green, and yellow curves, respectively) and b) the ODR profile sizes versus beam size for two different impact parameters (50 μm-blue curve; 100 μm-red curve). The vertical beam size was fixed at 10 μm, and the wavelength was 0.8 μm.

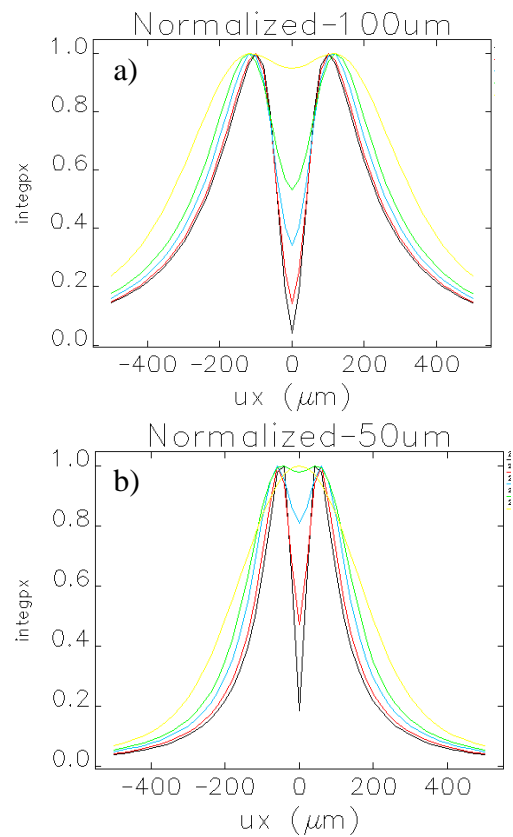

Figure 5: Numerical simulations of the ODR horizontally polarized component changes with horizontal beam size for two different impact parameters a) 100 μm and b) 50 μm. The input horizontal sizes are 10, 20, 35, 50, and 100 μm with the corresponding ODR profile curves in black, red, blue, green, and yellow, respectively. The vertical beam size was fixed at 10 μm, and the wavelength was 0.8 μm.

## SUMMARY

In summary, the feasibility of monitoring the 10- to 20-μm regime beam sizes with ODR at FACET has been evaluated by comparing to previous tests and by modeling. The prospects are sufficiently encouraging so that experiments are planned in the coming year as FACET is commissioned.

## ACKNOWLEDGEMENTS

The first author acknowledges the support of M. Wendt and M. Church of the Accelerator Division at Fermilab.